\documentclass[aps,10pt,twocolumn,showpacs,showkeys,nofootinbib]{revtex4-1}

\usepackage{amsmath}
\usepackage[pdftex]{graphicx}
\usepackage{enumerate}

\newcommand{\be}{\begin{equation}}
\newcommand{\ee}{\end{equation}}
\newcommand{\ba}{\begin{eqnarray}}
\newcommand{\ea}{\end{eqnarray}}

\def\g{\gamma}
\def\h{\eta}

\def\k{\kappa}
\def\l{\lambda}
\def\m{\mu}

\def\r{\rho}

\def\D{\Delta}

\newcommand{\ov}{\overline}

\newcommand{\aand}{\;\;\;\mbox{and}\;\;\;}
\newcommand{\pa}{\partial}

\def\I{\leavevmode\hbox{\small1\kern-3.8pt\norfeynmpmalsize1}}

\begin{document}
\title{Ostwald ripening for gas bubbles and decompression illness: phenomenological aspects in diving}

\begin{abstract}

The Ostwald ripening phenomenon for gas bubbles in a liquid consists mainly in gas transfer from smaller bubbles to larger bubbles. An experiment was carried out in which the Ostwald ripening for air bubbles, in a liquid fluid with some rheological parameters of the human blood, is reproduced. There it has been measured time evolution of bubbles mean radius, number of bubbles and radius size distribution, where the initial bubbles radii normalized distribution behaves like a Tsallis ($q$-Weibull) distribution. One of the main results shows that, while the number of bubbles decreases in time the bubbles mean radius increases, therefore smaller bubbles disappear whereas the, potentially dangerous for the diver, larger bubbles grow up. Consequently, it is presumed that such a bubble broadening effect could contribute, even minimally, to decompression illness: decompression sickness and arterial gas embolism. This conjecture is reinforced by the preliminary results of Ostwald broadening to RGBM (Reduced Gradient Bubble Model) decompression schedules for a closed circuit rebreather (CCR) dive to 420fsw (128m) with 21/79 Heliox gas mixture.





\end{abstract}

\author{O.M. Del Cima}
\email{oswaldo.delcima@ufv.br}

\author{C.M. Rocha}
\email{cesar.rocha@ufv.br}

\author{A.V.N.C. Teixeira}
\email{alvaro@ufv.br}

\affiliation{Universidade Federal de Vi\c cosa (UFV),\\
Departamento de F\'\i sica - Campus Universit\'ario,\\
Avenida Peter Henry Rolfs s/n - 36570-900 - Vi\c cosa - MG - Brazil.}

\author{B.R. Wienke}
\email{brw@lanl.gov}

\affiliation{Los Alamos National Laboratory (LANL),\\
Program Manager Computational Physics,  \\
C $\&$\,C Dive Team Ldr, \\
Los Alamos - NM 87545 - New Mexico - USA}


\maketitle

\section{Introduction}
The formation of gas bubbles \cite{diving_physics1,diving_physics2,diving_physics3} -- due to nucleation (homogeneous or heterogeneous) and tribonucleation, and their evolution (expansion or contraction), owing to decompression or compression, diffusion, counterdiffusion, coalescence and Ostwald ripening -- in the blood and tissues of the human body can give rise to the decompression illness (DCI) \cite{dci1,dci2,dci3}: decompression sickness (DCS) and arterial gas embolism (AGE). 

The Ostwald ripening mechanism consists in gas transfer from smaller bubbles to larger bubbles by diffusion in the liquid medium, consequently, the radii of larger bubbles increase at the expenses of decreasing radii of the smaller ones. 
It shall be presented the results of experiment and finite element simulation which the Ostwald ripening is investigated for the case of gas (air) bubbles in a liquid fluid with some rheological parameters of the human blood. There, it has been measured and analysed the time evolution of the bubbles mean radius, the number of bubbles and the radius size (frequency) distribution. At a fixed ambient pressure, namely, at the same ``depth'', one of the main experimental results has been undoubtedly shown that, while the number of bubbles decreases in time the bubbles mean radius increases, meaning that the smaller bubbles disappear whereas the larger (potentially dangerous) bubbles grow up. This phenomenon may reveal a contribution of the Ostwald ripening effect to the decompression illness during and after diving, suggesting, therefore, a deeper theoretical and experimental investigation. Beyond that, if the Ostwald ripening shows up as an important physiological effect to decompression sickness, its implementation to the RGBM (Reduced Gradient Bubble Model) \cite{rgbm1,rgbm2,rgbm3,rgbm4,rgbm5,rgbm6,rgbm7,blood} for further diving tests should be pursued.

The outline of this work is as follows. The processes of bubble evolution are briefly summarized in Subsection \ref{evolution}. The physiological consequences of bubble evolution are introduced in Subsection \ref{dcs1&2}. Section \ref{ostwald} considers essential basic theoretical aspects of the Ostwald ripening phenomenon for gas bubbles in a liquid fluid. The experiment is introduced in Section \ref{experiment}, wherein the experimental results of the Ostwald ripening effect among air bubbles in a liquid rheologically similar to the human blood are analysed, and the empirical model for the time evolution of bubbles mean radius, number of bubbles and radius size distribution are proposed, moreover it is verified that the bubbles initial radii normalized distribution behaves as a Tsallis ($q$-Weibull) distribution \cite{tsallis1,tsallis2}. Furthermore, in this section, preliminary results of Ostwald ripening (broadening effect) contribution to RGBM for a Closed Circuit Rebreather (CCR) dive to 420fsw (128m) with 21/79 Heliox (Heliair) gas mixture are presented and discussed. The conclusions and perspectives are left to Section \ref{conclusions}.  

\section{Bubbles evolution}
\subsection{Mechanisms of bubbles evolution} 
\label{evolution}
{\it Decompression or compression} -- 
Decompression (compression) is the process of decreasing (increasing) of the ambient pressure. 

{\it Coalescence} -- Coalescence is the fusion process of two or more bubbles. 

{\it Diffusion} -- Diffusion is the flow of substance (atoms or molecules) from regions of higher concentration to regions of lower concentration. 

{\it Isobaric counterdiffusion} -- Isobaric counterdiffusion is the diffusion of different gases into and out of tissues while under a constant ambient pressure \cite{lambertsen-idicula1,lambertsen-idicula2,wienke1,wienke2}. When in diving, with multiple inert gases, and performing an isobaric gas mix switch, the inert components of the initial mix breathed by the diver begin to off-­gas the tissues, whereas the inert components of the second mix begin to in-­gas the tissues. ``There is no change in pressure and the gases are moving in opposite directions, this is called {\it isobaric counterdiffusion}'' \cite{taylor}.

\begin{figure}[h!]
\centering
\setlength{\unitlength}{1,0mm}
\includegraphics[width=6.7cm,height=2.65cm]{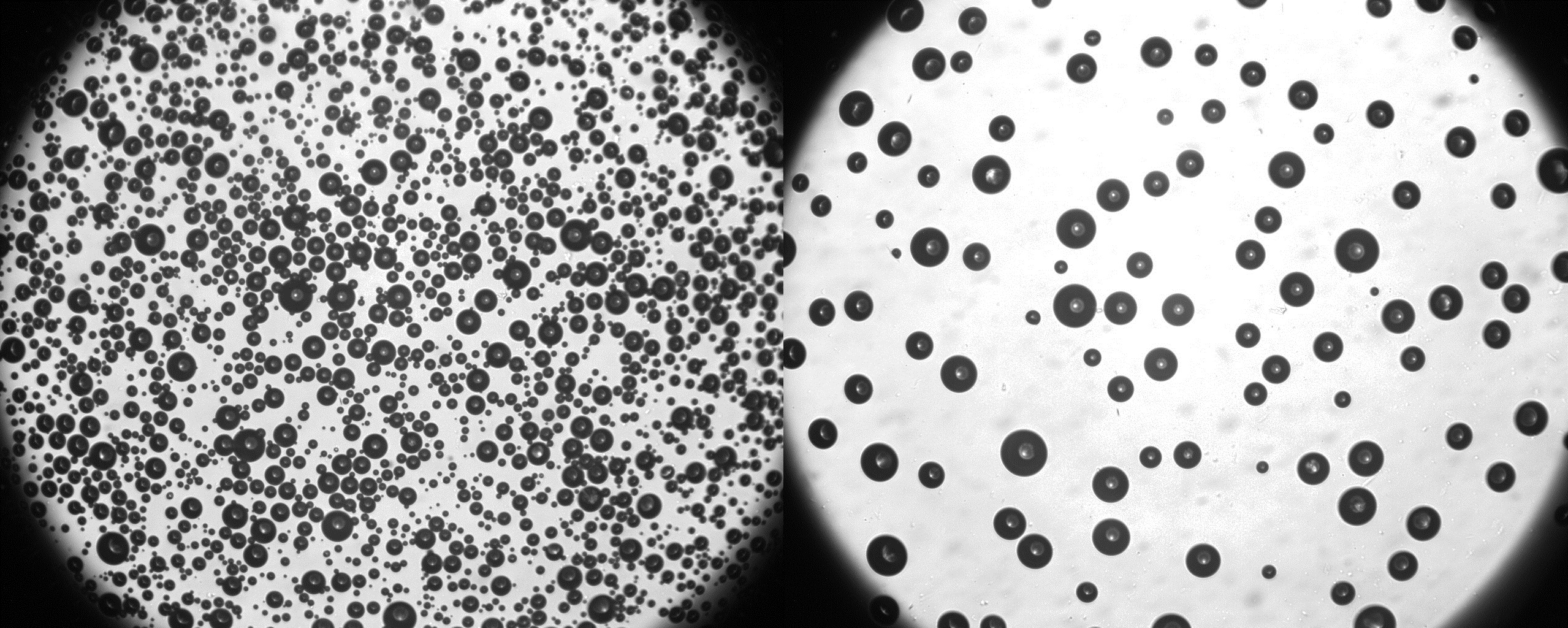}
\caption[]{Ostwald ripening: gas bubbles in a liquid fluid.}
\label{ostwald-gas-bubbles}
\end{figure}

{\it Ostwald ripening} -- In 1896, Wilhelm Ostwald has verified a phenomenon that small crystals (sol particles) dissolve and redeposit onto larger crystals (sol particles) \cite{ostwald}, which is called Ostwald ripening.  Later on, by means of the experiments presented and discussed in this work, it has been already observed the Ostwald ripening among gas bubbles in a liquid fluids, namely, gas transfer from smaller bubbles to larger bubbles (FIG. \ref{ostwald-gas-bubbles}).    

{\it Bubble evolution effects in deco models} -- Up to now, mixed phase decompression ({\it deco}) models \cite{deco1,deco2,rgbm2,deco4,deco5} -- reduced gradient bubble model, thermodynamic model, varying permeability model and tissue bubble diffusion model -- take into account, decompression/compression, diffusion and isobaric-counterdiffusion, as bubble evolution effects. Nevertheless, it remains to include within {\it deco} models the effects of coalescence and Ostwald ripening.

\subsection{Decompression sickness}
\label{dcs1&2}
One of the hazards that divers, astronauts, aviators and compressed air workers, are subjected while under hyperbaric or hypobaric conditions, submitted to compression and decompression, is the decompression illness (DCI) \cite{dci1,dci2,dci3}: decompression sickness (DCS) and arterial gas embolism (AGE). These gas bubbles injuries, that trigger, for instance, the decompression sickness, are due to the formation and evolution of intravascular and extravascular gas (N$_2$, He, O$_2$, CO$_2$, H$_2$O vapour) bubbles. The collective insult of the gas bubbles to the body shall produce primary effects to the tissues which are directly insulted, subsequently, the secondary effects can jeopardize the function of a wide range of tissues, compromising therefore body's health, may even lead to its death. Decompression sickness is recognized and classified by means of the signs and the symptoms exhibited by the body: Type 1  and Type 2. Type 1 DCS are usually characterized by mild cutaneous or skin symptoms, and musculoskeletal pain. Type 2 DCS symptoms are more severe, and they are typically split in three categories: cardiopulmonary, inner ear and neurological. 

\section{Ostwald ripening: gas bubbles in a liquid fluid}
\label{ostwald}
Smaller bubbles might {\it feed} larger bubbles -- the phenomenon of Ostwald ripening is ought to gas transfer from smaller bubbles to larger bubbles by diffusion in the liquid medium, provoking the radii increasing of larger bubbles at the expenses of decreasing radii of the smaller ones. 

The behaviour of a spherical gas bubble \cite{deGennes} at rest in a liquid fluid can be partially described by the Young-Laplace law: 
\be
\D p = p_{\rm bubble} - p_{\rm bulk} = \frac{2\g}{r}~, \label{young-laplace}
\ee
where $r$ is the bubble radius, $\g$ the surface tension and, $p_{\rm bubble}$ and $p_{\rm bulk}$ are the (gas) pressure inside and the (liquid) pressure outside the bubble, respectively. From Eq.(\ref{young-laplace}) it is straightforwardly verified that the gas bubble inner pressure ($p_{\rm bubble}$) is always greater than the outer pressure ($p_{\rm bulk}$), moreover, the smaller the bubble radius ($r$) the greater the pressure inside ($p_{\rm bubble}$) the bubble for a fixed ambient (outside) pressure ($p_{\rm bulk}$). Beyond the Young-Laplace law, the radius bubble kinetics is dictated also by:
\begin{enumerate}[(i)]
\item Fick's second law (diffusion):
\be
\frac{\pa c}{\pa t}=D\nabla^2c ~, \label{fick}
\ee
where $D$ is the gas/liquid diffusion coefficient and $c$ the gas concentration in the liquid phase; 
\item Henry's law (solubility):
\be
c_{\operatorname{b-edge}} = H p_{\rm bubble} ~, \label{henry}
\ee  
with $H$ being the gas/liquid Henry solubility and $c_{\operatorname{b-edge}}$ the gas concentration at bubble-edge (gas-liquid interface).     
\end{enumerate}

Finally, taking into consideration the laws of Young-Laplace (\ref{young-laplace}), Fick (\ref{fick}) and Henry (\ref{henry}), with $c_{\operatorname{e-bulk}}$ being the bubble edge-bulk gas concentration, it can be concluded that the bubble diminishes (bubble releasing gas to liquid) while $c_{\operatorname{b-edge}}>c_{\operatorname{e-bulk}}$, whereas the bubble expands (liquid releasing gas to bubble) while $c_{\operatorname{b-edge}}<c_{\operatorname{e-bulk}}$. Consequently, when $c_{\operatorname{b-edge}}=c_{\operatorname{e-bulk}}$ a single bubble lays in equilibrium since its radius ($r$) is equal to the critical radius ($\epsilon$) -- called bubble excitation radius in RGBM -- $r=\epsilon$, meanwhile if $r<\epsilon$ the bubble collapses otherwise if $r>\epsilon$ the bubble grows.

\begin{figure}[h!]
\centering
\setlength{\unitlength}{1,0mm}
\includegraphics[width=5.2cm,height=2.65cm]{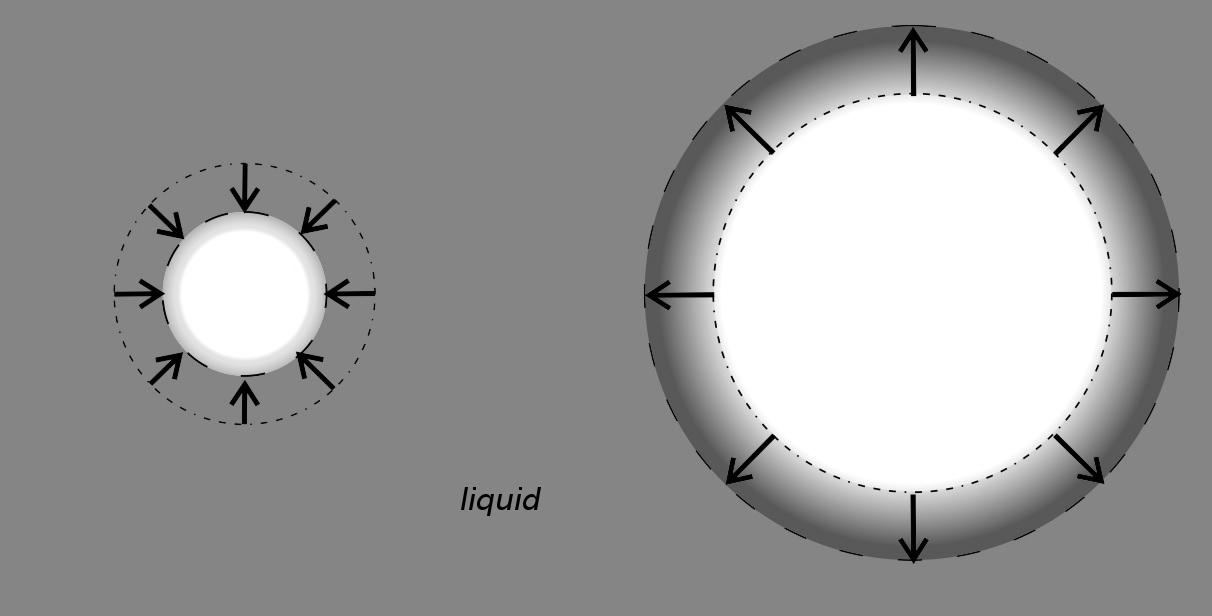}
\caption[]{Two gas bubbles with different radii in a liquid.}
\label{two-bubbles}
\end{figure}

Now, what shall happen if there are two gas bubbles with different radii into the liquid (FIG.\ref{two-bubbles})? Taking into account the Young-Laplace equation (\ref{young-laplace}) for the two bubbles, it follows that:  
\be
p_{\rm bubble} = p_{\rm bulk} + \frac{2\g}{r} \aand P_{\rm bubble} = p_{\rm bulk} + \frac{2\g}{R}~, \label{young-laplace-two}
\ee
where $r$ and $R$ are de radii of the smaller and greater bubbles ($r<R$), respectively, the ambient (liquid) pressure is $p_{\rm bulk}$, $p_{\rm bubble}$ is the inner pressure of the smaller bubble, and $P_{\rm bubble}$ the inner pressure of the greater one. The Young-Laplace equations displayed in (\ref{young-laplace-two})  straightforwardly indicates that $r<R$ implies $p_{\rm bubble}>P_{\rm bubble}$, this fact together with the last paragraph discussion and due to the edge-bulk gas concentration gradient among the two bubbles, assuming a fixed ambient (liquid) pressure which is the experiment condition (case of a diver at a fixed depth, where $p_{\rm bulk}$ is constant), the Ostwald ripening -- gas flow from the smaller to the larger bubble -- might be observed. Complex systems of gas bubbles in liquid (FIG.\ref{ostwald-gas-bubbles}), as the experiment (FIG.\ref{bubble-chamber}) presented and discussed here, or into the non-newtonian fluids and tissues of living organisms exhibit Ostwald broadening. Nevertheless, in order to a better comprehension of possible effects owing to Ostwald ripening in high complex systems like the human body, it is essential to go further into details on bubbles broadening in less complex, and more controlled, systems as the one analysed in this work.    

Huge amount of gas (N$_2$, He, O$_2$, CO$_2$, H$_2$O vapour) bubbles are dispersed throughout the body (inside tissues and blood) of a diver, astronaut, aviator or compressed air worker, whose radii vary from $10^{-1}\m$m to $10^{2}\m$m. The evolution of these gas bubbles is quite complicated since it involves altogether, compression/decompression, diffusion, isobaric counterdiffusion, coalescence, Ostwald ripening (broadening effect) and besides other more complex phenomena, therefore, experimental, theoretical and computational attempts to investigate, understand and describe such kind of complex system are herculean tasks. 

\section{Ostwald ripening: the experiment}
\label{experiment}
The main purpose of this work is to present some preliminary experimental results, and propose an empirical model, on the Ostwald ripening for gas bubbles \cite{bubble-group-article} in a liquid with some rheological parameters (density, surface tension and viscosity) of the human blood \cite{blood}, and discuss its possible implications to the risk of decompression sickness. In addition to that, the Ostwald broadening empirical model introduced here was applied for the first time to RGBM for a CCR dive to 128m with Heliox 21/79 and the dive decompression schedule results are displayed following in TAB.\ref{table1}. 

\begin{figure}[h!]
\centering
\setlength{\unitlength}{1,0mm}
\includegraphics[width=5.2cm,height=2.65cm]{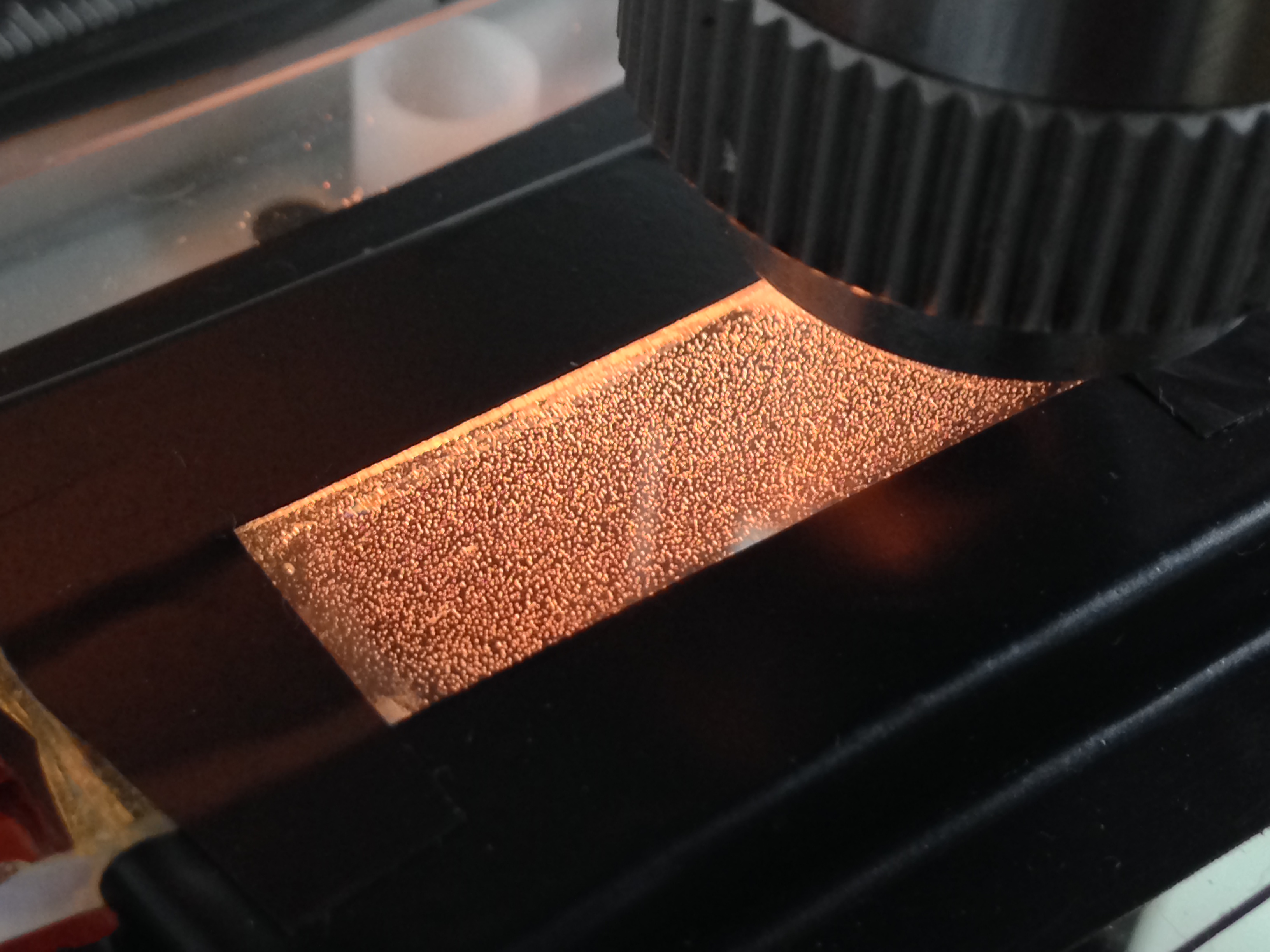}
\caption[]{Air bubbles confined in the chamber.}
\label{bubble-chamber}
\end{figure}

The physical quantities adopted, in order to describe the time evolution of the whole system consisting of air bubbles in a liquid solution confined in a (bubble) chamber, are: mean bubble radius ($\ov{R}(t)$), number of bubbles ($N(t)$), radii (frequency) distribution ($f(R,t)$) and radii normalized (probability) distribution ($p(R,t)$). The experiment apparatus includes an optical microscope (10x lense), a B\&W camera coupled to the microscope and connected to a computer, a bubble chamber attached to a displacement table controlled by the computer, moreover, the air bubbles in the liquid solution, prior to injection into the bubble chamber (FIG.\ref{bubble-chamber}), are produced by cavitation. 

\begin{figure}[h!]
\centering
\setlength{\unitlength}{1,0mm}
\includegraphics[width=5.2cm,height=4.2cm]{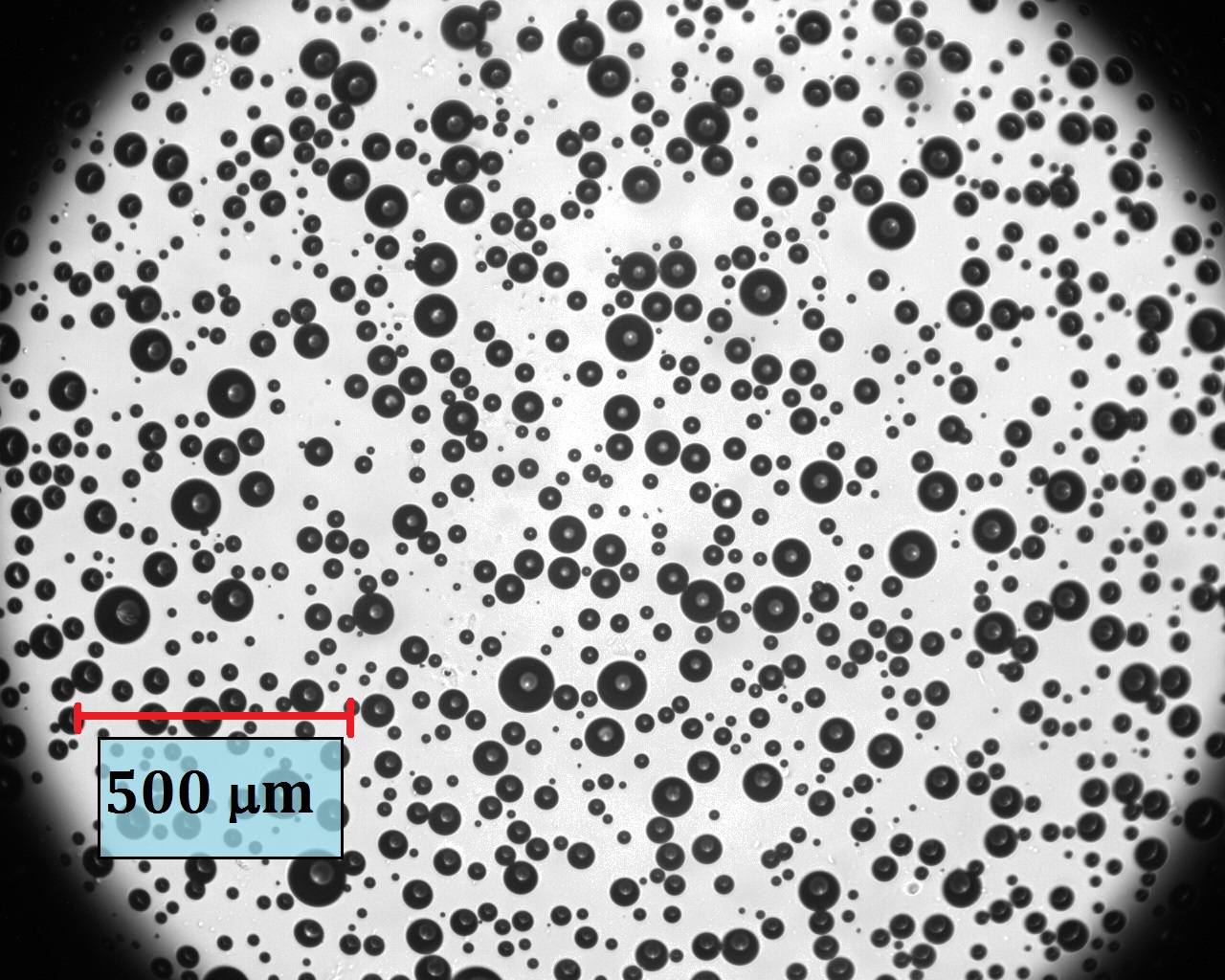}
\caption[]{Air bubbles: microscope focus.}
\label{microscope-focus}
\end{figure}

The liquid solution (v/v) used in the experiment contains 75\% of glycerol and 25\% of deionized water, exhibiting the following measured rheological parameters:
\ba
& {\rm density} \longrightarrow \r_{\rm exp} = (1,17 \pm 0,01)~{\rm g}\,{\rm cm}^{-3}~,\nonumber\\
& {\rm surface~tension} \longrightarrow \g_{\rm exp} = (65,3 \pm 0,01)~{\rm mN}\,{\rm m}^{-1}~,\nonumber\\
& {\rm viscosity} \longrightarrow \h_{\rm exp} = (34,530 \pm 0,002)~{\rm mPa}\,{\rm s}~,\label{exp}
\ea 
at $25^{\circ}{\rm C}$ room temperature. It shall be noticed that, in order to set apart (experimentally) the Ostwald ripening from other effects, namely, potential coalescence among air bubbles and also their dislocations (FIG.\ref{microscope-focus}), the liquid solution viscosity, $\h_{\rm exp}$ (\ref{exp}), had to be fixed greater than the mean blood viscosity, $\h_{\rm blood}$ \cite{blood}:
\ba
& {\rm density} \longrightarrow 1,00 \leq \r_{\rm blood} \leq 1,15~({\rm g}\,{\rm cm}^{-3})~,\nonumber\\
& {\rm surface~tension} \longrightarrow 15 \leq \g_{\rm blood} \leq 80~({\rm mN}\,{\rm m}^{-1})~,\nonumber\\
& {\rm viscosity} \longrightarrow 1,00 \leq \h_{\rm blood} \leq 4,00~({\rm mPa}\,{\rm s})~.\label{blood}
\ea 

\begin{figure}[h!]
\centering
\setlength{\unitlength}{1,0mm}
\includegraphics[width=9.0cm,height=6.0cm]{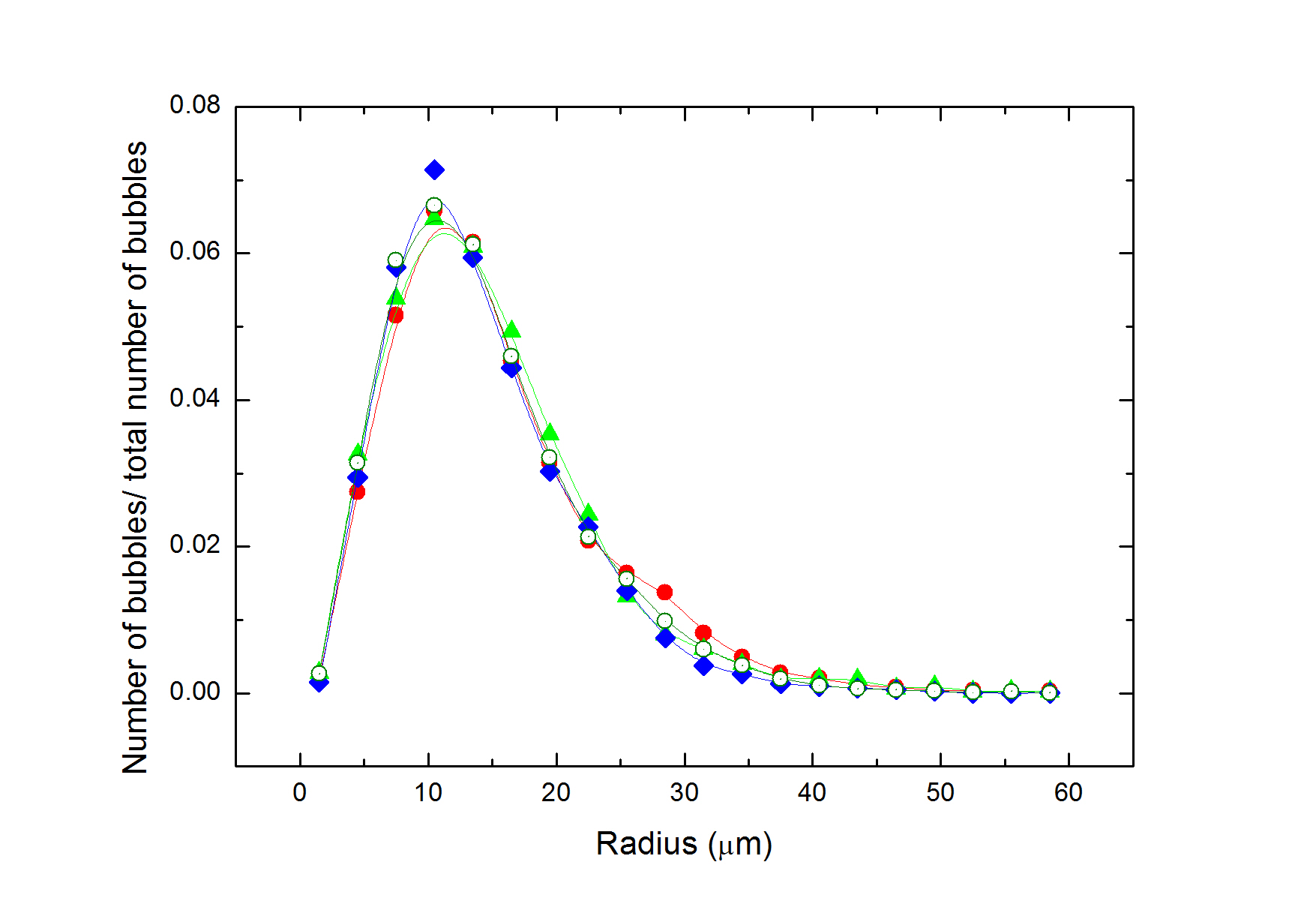}
\caption[]{Reproducibility: the initial radii normalized distribution ($p(R,0)$) for the four samples}
\label{four-initial-distributions}
\end{figure}

The experiment ran for four samples of air bubbles in the liquid solution. The initial radii normalized distribution ($p(R,0)$) -- defined by the ratio among the number of bubbles with radius $R$ (the radii distribution $f(R,0)$) and the total number of bubbles ($N(0)\sim 10^4$) at instant zero ($0$h) -- of the four samples analysed, validates the reproducibility of the experiment (FIG.\ref{four-initial-distributions}). 

\begin{figure}[t]
\centering
\setlength{\unitlength}{1,0mm}
\includegraphics[width=9.0cm,height=6.0cm]{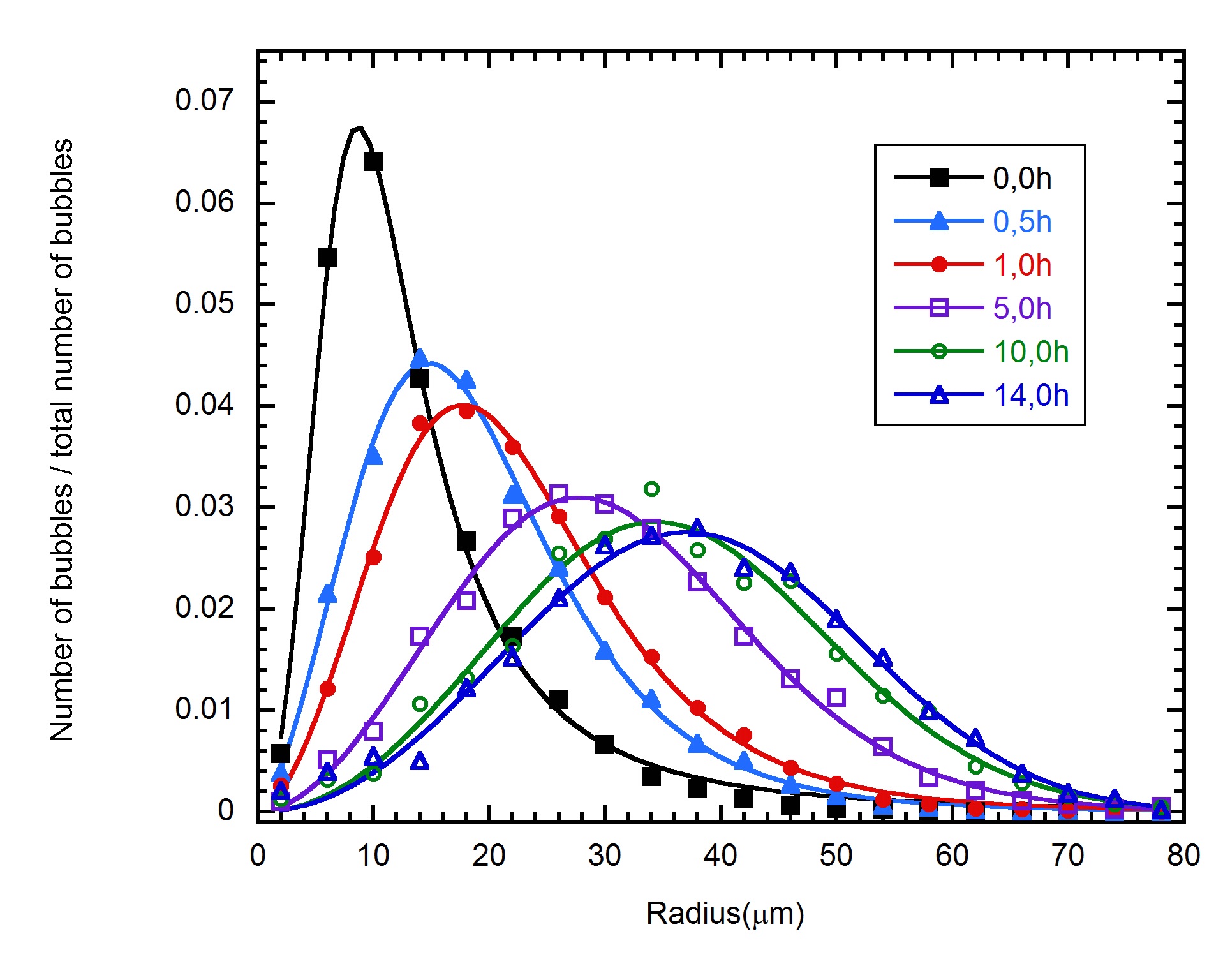}
\caption[]{The radii normalized (probability) distribution ($p(R,t)$): $0$; $0,5$; $1$; $5$; $10$ and $14$h.}
\label{normalized-distribution}
\end{figure}

The radii (frequency) distribution as a function of time is acquired analysing \cite{imagej} the images taken by the B\&W camera (coupled to the microscope) at different zones of the bubble chamber reached by means of the displacement table controlled by the computer. The acquisition of the images by the B\&W camera and the subsequent bubbles radii measurement, the radii (frequency) distribution and the radii normalized (probability) distribution (FIG.\ref{normalized-distribution}) were obtained at the time instants: $0$; $0,5$; $1$; $5$; $10$ and $14$h. In what concerns the bubbles time evolution, it can be deduced from FIG.\ref{normalized-distribution} that the number of bubbles decreases whereas the mean bubble radius increases. 
The initial radii normalized distribution ($p(R,0)$) behaves as a Tsallis ($q$-Weibull) distribution \cite{tsallis1,tsallis2}:
\be
p(R,0) = (2-q) \frac{\k}{\l} \left(\frac{R}{\l}\right)^{\k-1} 
\left[ 1-(1-q)\left(\frac{R}{\l}\right)^\k \right]^{\frac{1}{1-q}} ~, \label{tsallis}
\ee
where the fit fixes $q=1,46$, $\k=2,34$ and $\l=12,43\m{\rm m}$ at instant zero ($0$h).

\begin{figure}[t]
\centering
\setlength{\unitlength}{1,0mm}
\includegraphics[width=8.1cm,height=5.4cm]{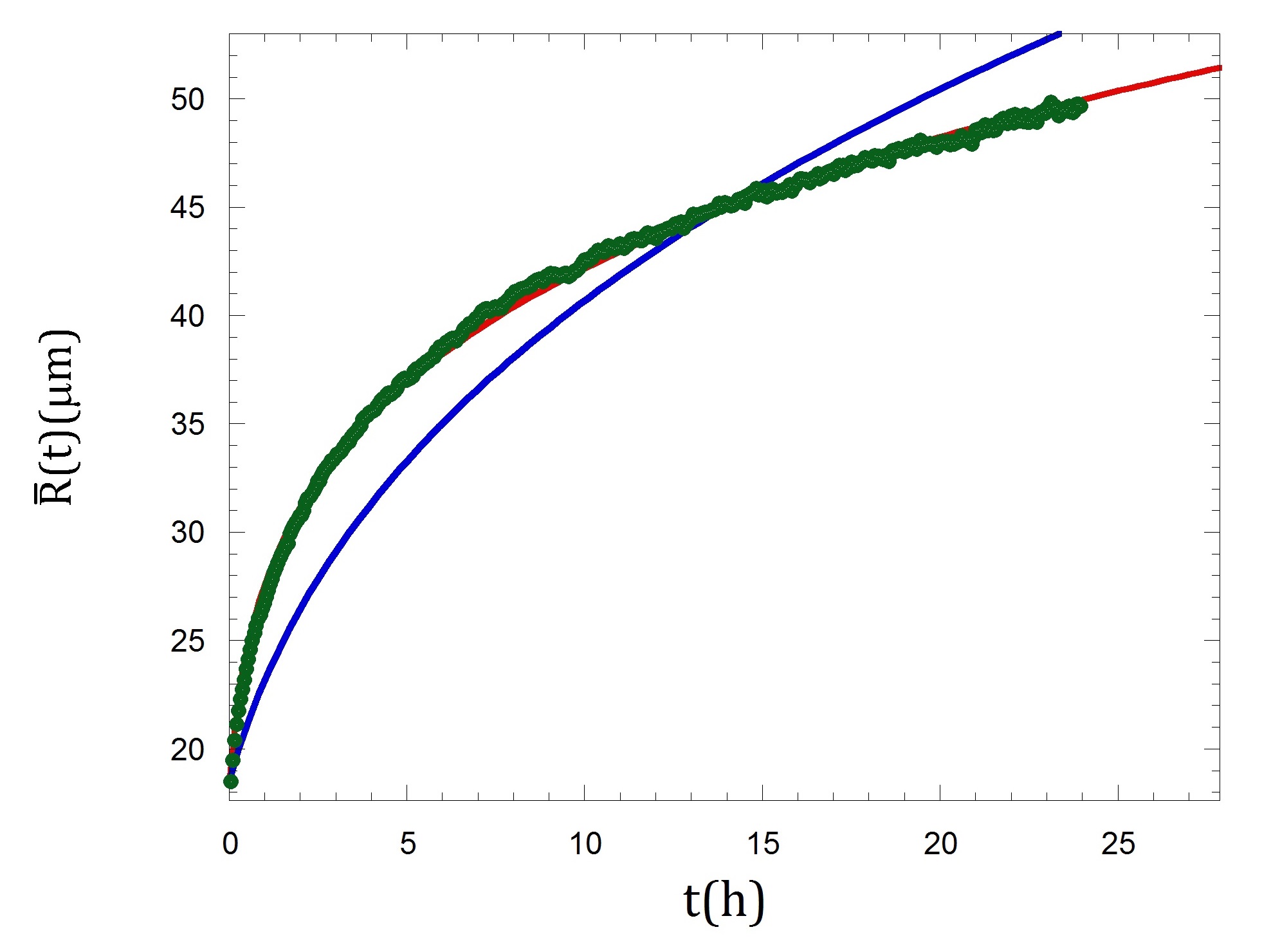}
\caption[]{The mean bubble radius: the experimental data fit ($\ov{R}(t)$ -- red line) and the LSW theory prediction ($\ov{R}_{\rm LSW}(t)$ -- blue line) assuming the same experimental initial condition, the initial mean bubble radius 
$\ov{R}(0)=18,42\m$m.}
\label{mean-radius}
\end{figure}

The experimental results for the mean bubble radius ($\ov{R}(t)$) show that it increases monotonically in time, which can be straightforwardly concluded from FIG.\ref{mean-radius}, wherein, together with the experimental curve fit, it is also sketched a curve for the mean bubble radius ($\ov{R}_{\rm LSW}(t)$) if the bubbles dynamics was dictated by the LSW (Lifshitz-Slyozov-Wagner) theory \cite{lsw1,lsw2}. The mean bubble radius ($\ov{R}_{\rm LSW}(t)$) in LSW theory is given by:
\ba
&\ov{R}_{\rm LSW}(t) = \left[\ov{R}^3(0) + Kt\right]^{\frac{1}{3}} ~,~~ \ov{R}(0)=18,42\m{\rm m} \nonumber\\ 
&\aand K=6,1\times 10^3{\rm m}^3{\rm s}^{-1} ~, \nonumber 
\ea
where $K$ depends on the temperature, surface tension, diffusion coefficient, gas solubility and gas molar volume. Therefore, it can be verified from FIG.\ref{mean-radius} (blue line) that the LSW theory does not describe a system of gas bubbles in a liquid, which should be expected since such a system does not satisfy the primary premises assumed by the LSW theory, namely, vanishing volume fraction, bubbles radii much smaller than the typical distance between the nearest neighbours and instantaneous homogeneity of gas concentration throughout the bulk. Nevertheless, bearing in mind the experimental data acquired and modelling the mean bubble radius ($\ov{R}(t)$) as below:
\ba
&\ov{R}(t) = \left[\ov{R}^{\frac{1}{\chi}}(0) + Kt\right]^{\chi} ~, \label{rbar-time}\\
&\ov{R}(0)=18,42\m{\rm m} ~,~~ K=2,0\times 10^7{\rm m}^{\frac{1}{\chi}}{\rm s}^{-1} \nonumber\\ 
&\aand \chi=0,1956 ~, \nonumber 
\ea
from FIG.\ref{mean-radius} (red line) it follows that the empirical model (\ref{rbar-time}) proposed fits the experimental data.  

\begin{figure}[t]
\centering
\setlength{\unitlength}{1,0mm}
\includegraphics[width=8.1cm,height=5.4cm]{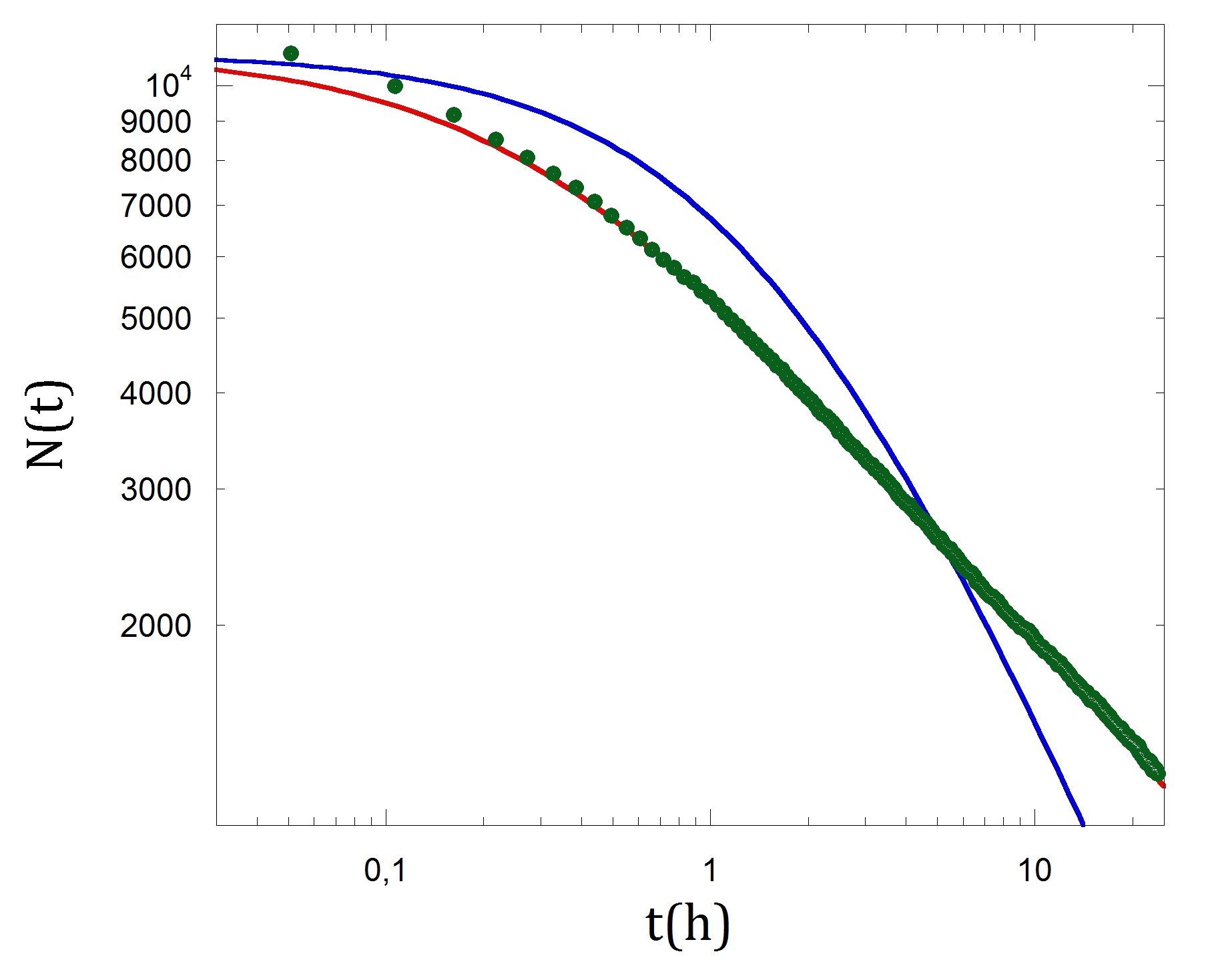}
\caption[]{The number of bubbles: the experimental data fit ($N(t)$ -- red line) and the LSW theory prediction ($N_{\rm LSW}(t)$ -- blue line) assuming the same experimental initial condition, the mean bubble radius, $\ov{R}(0)=18,42\m$m, and the number of bubbles, $N(0)=1,8\times 10^4$.}
\label{number}
\end{figure}

On the other way around, the experiment shows that, while the mean bubble radius ($\ov{R}(t)$) increases in time, the number of bubbles ($N(t)$) decreases monotonically as displayed in FIG.\ref{number}, where it can also be seen that the number of bubbles ($N_{\rm LSW}(t)$) proposed by the LSW theory: 
\ba
&N_{\rm LSW}(t) = N(0) \displaystyle\frac{\ov{R}^3(0)}{\ov{R}^3(0) + Kt} ~,~~ \ov{R}(0)=18,42\m{\rm m}~, \nonumber\\ 
&N(0)=1,8\times 10^4 \aand K=6,1\times 10^3{\rm m}^3{\rm s}^{-1} ~, \nonumber 
\ea
does not fit (blue line) the experimental data. However, by considering the experimental data, and modelling the number of bubbles ($N(t)$) as follows:
\ba
&{N}(t) = N(0) \displaystyle\frac{\ov{R}^{\frac{\l}{\chi}}(0)}{\left[\ov{R}^{\frac{1}{\chi}}(0) + Kt\right]^{\l}} ~, \label{n-time} \\ 
&\ov{R}(0)=18,42\m{\rm m} ~,~~ K=2,0\times 10^7{\rm m}^{\frac{1}{\chi}}{\rm s}^{-1} ~,\nonumber\\
&\chi=0,1956 \aand \l=0,48 ~, \nonumber 
\ea
it can be verified that the empirical model (\ref{n-time}) fits the experimental data, FIG.\ref{number} (red line). 

In summary, the experiment realized -- upon a system of air bubbles in a liquid fluid with some rheological parameters of the human blood -- shows that the mean bubble radius ($\ov{R}(t)$) increases (FIG.\ref{mean-radius}) in time whereas the number of bubbles ($N(t)$) decreases (FIG.\ref{number}). Moreover, 
it is verified straightforwardly that the smaller bubbles disappear whereas the larger bubbles, which trigger the decompression sickness, grow up. Based on the experimental analysis of the Ostwald broadening phenomenon, it was proposed an empirical model for the mean bubble radius (\ref{rbar-time}) and the number of bubbles (\ref{n-time}), which properly describes the time evolution of the system. Taking into consideration the Ostwald ripening empirical model proposed here, it is interesting to probe its implementation in the Reduced Gradient Bubble Model (RGBM) in such a manner to compute decompression sickness risks and develop dive tables for further diving tests. Some preliminary results follow.

In RGBM, permissible supersaturations controlling diver ascents depend on bubble excitation radii ($\epsilon$) defined as the minimum bubble radii separating growing from collapsing bubbles. With broadening, these radii are increased in magnitude rendering permissible supersaturations smaller. Smaller permissible supersaturations decrease no-decompression time limits (NDLs) and increase decompression times. The increases in excitation radii ($\epsilon$) scale multiplicatively with the ratio of broadened over unbroadened mean bubble distribution radii, $r$. The rheological (tissue and blood) extensions of the transport coefficients, $K$, in the hydrocarbon mixture (75\% C$_3$H$_8$O$_3$ + 25\% H$_2$O) experiments scale as $(7,8)^{-1}K$ for this study. Mathematical details for the interested reader can be found in the literature \cite{wienke1,wienke2}. For this study, we use the results showed here though differences with the ongoing experiment \cite{bubble-group-article}. With these modifications to the RGBM, we cautiously make some hypothetical estimates from these laboratory experiment erring on the conservative side and assuming bubble broadening might occur in the body on the same time scales as in the laboratory. For meter manufacturers, diveware purveyors, modelers and table designers these modifications to RGBM (and USN, ZHL and VPM)) are easily made \cite{wienke1,wienke2} and are always conservative.

\begin{table}[t]
\begin{tabular}{ c c c }
       & ~{\it no }~ & ~{\it Ostwald}~   \\
       & ~{\it broadening}~ & ~{\it broadening}~   \\
 depth & time & time  \\
 (fsw) & (min)& (min) \\   
    &       &          \\
420 &  15,0 & 15,0  \\
340 &   0,0 &  0,0   \\
330 &   0,5 &  0,5   \\
320 &   0,5 &  0,5   \\
310 &   0,5 &  0,5   \\
300 &   0,5 &  0,5   \\
290 &   1,0 &  1,0   \\
280 &   1,0 &  1,0   \\
270 &   1,0 &  1,0   \\
260 &   1,0 &  1,0   \\
250 &   1,5 &  1,5   \\
240 &   1,5 &  1,5   \\
230 &   1,5 &  1,5   \\
220 &   1,5 &  1,5   \\
210 &   1,5 &  2,0   \\
200 &   2,0 &  2,0   \\
190 &   2,0 &  2,0   \\
180 &   2,0 &  2,0   \\
170 &   2,5 &  3,0   \\
160 &   3,0 &  3,0   \\
150 &   3,0 &  3,0   \\
140 &   4,0 &  4,0   \\
130 &   5,0 &  5,0   \\
120 &   5,0 &  5,0   \\
110 &   5,5 &  5,5   \\
100 &   6,5 &  7,5   \\
 90 &   7,0 &  7,5   \\
 80 &   7,0 &  9,0   \\
 70 &  10,0 & 10,5   \\
 60 &  10,0 & 11,5   \\
 50 &  13,5 & 14,5   \\
 40 &  15,0 & 17,5   \\
 30 &  18,0 & 21,5   \\
 20 &  24,0 & 26,5   \\
 10 &  30,5 & 35,5   \\
    & 219,5 & 228,5  \\
\end{tabular}
\caption[]{Decompression schedules for 420fsw (128m) CCR dive and bubble broadening.}\label{table1}
\end{table}

The TABLE \ref{table1} contrasts broadening for a Heliair (21/79 Heliox or EAH21) CCR (closed circuit reabreather) dive to 420fsw (128m) for 15min with setpoint 1,1atm and broadening time scale bottom time. This is not an easy dive by any means, technical or otherwise. If broadening time scales include decompression time, effects are larger on the order of 12\%. Extreme dives like this are not recorded often but reports from the field anecdotally suggest they are being made more commonly these days on rebreather systems. Beyond 4h broadening time scales hypothetical effects disappear completely. This also is not a nominal dive and hypothetical effects of broadening are more pronounced here. It is in the LANL DB too having been performed by the C\&C Team off Dry Tortugas. 

\section{Conclusions and perspectives}
\label{conclusions}
The Ostwald ripening, a broadening phenomenon due to gas diffusion among bubbles, which results that larger bubbles are fed by the smaller, are reproduced and described experimentally. The experiment consisted of air bubbles into a chamber filled by a liquid solution with some human blood-like rheological parameters, density and surface tension. 

The experiment which ran at $25^{\circ}{\rm C}$ under normobaric ambient pressure showed that the number of bubbles -- the initial number of bubbles was of order $10^4$ -- decreases in time while the mean bubble radius  increases. Additionally, from the initial experimental data it has been already verified that the radii normalized distribution (\ref{tsallis}) is a Tsallis ($q$-Weibull) distribution \cite{tsallis1,tsallis2}. It is proposed, analysing the experimental results, an empirical model for the Ostwald ripening by describing the time evolution of the number of bubbles (\ref{n-time}) and the mean bubble radius (\ref{rbar-time}). It shall be stressed that one of the main results shown, even at a constant ambient pressure (at the same diver ``depth''), a decreasing number of bubbles in time while the bubbles mean radius increases, consequently, the smaller bubbles disappear whereas the larger (potentially dangerous to divers) bubbles grow up, hence this might reveal a contribution of the Ostwald broadening to the decompression sickness risk during and after diving.

There are many perspectives and challenges to be pursued. From the experimental point of view, and for further computational simulation \cite{finite-element}, it is important to search for mechanisms to suppress or to promote the Ostwald ripening. Besides, the experiment shall be performed at typical human body temperatures, around $36,5\mbox{--}37,5^{\circ}{\rm C}$, for nitrogen bubbles into human plasma, also by adding to the plasma polystyrene microdisks, with diameters about $6\mbox{-}8\m{\rm m}$, simulating the red blood cells. The Ostwald ripening empirical model -- for the number of bubbles (\ref{n-time}) and the mean bubble radius (\ref{rbar-time}) in time -- has been implemented in the Reduced Gradient Bubble Model (RGBM) \cite{rgbm1,rgbm2,rgbm3,rgbm4,rgbm5,rgbm6,rgbm7,blood}, first to recreational air diving protocols, in order to obtain the ``new'' risk estimates for various NDLs (no-decompression time limits) and compare them to those of RGBM, as well as to those from other models, namely, ZHL (B\"uhlmann), USN (U.S. Navy) and VPM (Varying Permeability Model). Some simple results of bubble broadening for very deep decompression diving on CCRs with helium have been presented showing effects that increase decompression time with both depth and time. Overall the effects are small to moderate for nominal technical/recreational diving.      

\newpage

\begin{acknowledgments}
 
O.M.D.C and A.V.N.C.T thank C. Tsallis and M.L. Martins for discussions and suggestions. FAPEMIG, FUNARBE, CNPq and CAPES are acknowledged for invaluable financial help. O.M.D.C. and B.R.W. thank A. Oliveira, L. Notomi, T. O'Leary, NAUI Worldwide and NAUI Brasil.

\end{acknowledgments}

\bibliographystyle{apsrev4-1}
\bibliography{ensembles}

\end{document}